\documentclass[prd,aps,showpacs,nofootinbib,amsmath,amssymb]{revtex4}
\usepackage{amssymb,epsfig}
\usepackage[usenames,dvipsnames]{color}
\usepackage[bookmarks]{hyperref}
\usepackage[usenames,dvipsnames]{color}

\newtheorem{lemma}{Lemma}

\newcommand{\proof}{\noindent {\bf Proof. }}

\newcommand{\qed}{\hfill $\fbox{\hspace{0.3mm}}$ \vspace{.3cm}} %End of proof.

\newcommand{\Real}{\mathbb{R}}

\begin{document}
%%%%%%%%%%%%%%%%%%%%%%%%%%%%%%%

\title{Michel accretion of a polytropic fluid with adiabatic index $\gamma > 5/3$: 
Global flows versus homoclinic orbits}

\author{Eliana Chaverra$^{1,2}$, Patryk Mach$^3$, and Olivier Sarbach$^1$}
\affiliation{$^1$Instituto de F\'isica y Matem\'aticas,
Universidad Michoacana de San Nicol\'as de Hidalgo,
Edificio C-3, Ciudad Universitaria, 58040 Morelia, Michoac\'an, M\'exico,\\
$^2$Escuela Nacional de Estudios Superiores, Unidad Morelia, Universidad Nacional Aut\'onoma de M\'exico, Campus Morelia, C.P. 58190, Morelia, Michoac\'an, M\'exico,\\
$^3$Instytut Fizyki im.~Mariana Smoluchowskiego, Uniwersytet Jagiello\'{n}ski, {\L}ojasiewicza 11, 30-348 Krak\'{o}w, Poland}

\begin{abstract}
We analyze the properties of a polytropic fluid which is radially accreted into a Schwarzschild black hole. The case where the adiabatic index $\gamma$ lies in the range $1 < \gamma \leq 5/3$ has been treated in previous work. In this article we analyze the complementary range $5/3 < \gamma \leq 2$. To this purpose, the problem is cast into an appropriate Hamiltonian dynamical system whose phase flow is analyzed. While for $1 < \gamma \leq 5/3$ the solutions are always characterized by the presence of a unique critical saddle point, we show that when $5/3 < \gamma \leq 2$, an additional critical point might appear which is a center point. For the parametrization used in this paper we prove that whenever this additional critical point appears, there is a homoclinic orbit.
\end{abstract}

\date{\today}

% 04.20.-q: Classical GR
% 04.20.Ex: Initial value problem, existence and uniqueness of solutions
% 04.25.-g: Approximation methods; equations of motion
% 04.25.D-: Numerical relativity
% 04.25.Nx: Post-Newtonian approximation; perturbation theory; related approximations
% 04.40.-b: Self-gravitating systems, continuous media and classical fields in curved spacetime
% 04.70.-s: Physics of black holes
% 97.60.Lf: Astronomy: Late stage of star evolution: black holes
% 98.62.-g: Characteristics and properties of external galaxies and extragalactic objects
% 98.62.Mw: Infall, accretion, and accretion disks

\maketitle

%%%%%%%%%%%%%%%%%%%%%%%%%%%%%%%%%%%%%%%%%%%%%
\section{Introduction}
%%%%%%%%%%%%%%%%%%%%%%%%%%%%%%%%%%%%%%%%%%%%%

A spherically symmetric accretion model introduced by Bondi in~\cite{hB52} belongs to classical textbook models of theoretical astrophysics. Its general-relativistic version was proposed by Michel~\cite{fM72}, who considered a spherically symmetric, purely radial, stationary flow of perfect fluid in the Schwarzschild spacetime. Since then, different variants of this model were studied in numerous works that took into account the self-gravity of the fluid, radiation transfer, electric charge, cosmological constant, non-zero angular momentum, etc.\ (see for instance \cite{eM99, jKbKpMeMzS06, pMeM08, jKeMkRzS09, eMtR10, vDyE11, jKeM13, eBvDyE11, sGsKaRtD07, tDbC12, eTsMjM12, eTpTjM13, fLmGfG14}).

The standard way of parametrizing Bondi--Michel solutions is to prescribe the asymptotic values of the speed of sound and the density. By adopting such a parametrization one assumes {\em eo ipso} that the solution is global, i.e., it extends all the way from the horizon to infinite radii. Local solutions that cannot be extended to infinity were recently discovered in the cosmological context in~\cite{pMeMjK13, pMeM13, pM15, fF15} and, more surprisingly, in the standard Michel accretion (on the Schwarzschild background) of polytropic fluids with adiabatic indices $\gamma > 5/3$ \cite{Eliana-Master-thesis,eCoS12,eCoS15a}. Interestingly, these local solutions correspond to homoclinic orbits appearing in phase diagrams of the radial velocity vs.\ radius or the density vs.\ radius. Thus, from the mathematical point of view, the question of the existence of local vs.\ global solutions translates into the problem of the existence of homoclinic solutions in the appropriate phase diagrams.

In \cite{pM15} this issue was investigated in the case of the Bondi--Michel accretion in Schwarzschild--anti-de Sitter spacetime for polytropic and isothermal (linear) equations of state. It was shown that for polytropic equations of state, with a contribution in the specific enthalpy due to the rest-mass (baryonic) density, only local, homoclinic solutions are allowed. In contrast to that, for isothermal equations of state of the form $p = k e$, where $p$ is the pressure, $e$ is the energy density, and $0 < k \leq 1$ is a constant, global solutions do exist for $k \ge 1/3$, while only local, homoclinic solutions are allowed for $k < 1/3$. The limiting value $k = 1/3$ corresponds exactly to the simple photon gas model.

It is thus quite surprising to observe that in the classic case of a Schwarzschild black hole (vanishing cosmological constant) and polytropic fluids, both types of solutions are allowed for $\gamma > 5/3$. Unfortunately, it seems at present that it is difficult to formulate any simple and strict statement on the connection between the type of the equation of state and the existence of global or homoclinic solution. One should also note that the arguments given in \cite{pM15} are not as mathematically strict as one would expect them to be. For example, no general proof of the existence of homoclinic solutions is given. Their occurrence is rather observed on specific examples. In this paper, dealing with the relatively simple (yet probably the most important) case of accretion onto a Schwarzschild black hole, we are able to provide rigorous results concerning the existence of both homoclinic and global solutions, as well as the number and types of critical points on the appropriate phase diagrams. Although partial results hold for a quite general class of barotropic equations of state, we focus mainly on the polytropic Michel accretion with $5/3 < \gamma \le 2$. A preliminary analysis regarding the structure of the critical points in this model can be found in \cite{mB78}, where it is assumed that the fluid particles are nonrelativistic at infinity. This choice corresponds to $L^2$ slightly larger than one in our notation below, and it excludes the existence of homoclinic orbits.

The order of this paper is as follows. The equations governing the Michel accretion are collected in the next section. In Sec.\ \ref{sec_dyn_system} we discuss different possibilities for defining a Hamiltonian dynamical system corresponding to the equations of the Michel model. Section \ref{Sec:Crit} contains a characterization of the critical points, while in Sec.\ \ref{Sec:Global} we state our main results on the topology of the orbits. Section \ref{Sec:Conclusions} contains a brief summary and implications of our work.

%%%%%%%%%%%%%%%%%%%%%%%%%%%%%%%%%%%%%%%%%%%%%
\section{Michel accretion model}
%%%%%%%%%%%%%%%%%%%%%%%%%%%%%%%%%%%%%%%%%%%%%

In this section, we briefly review the relevant equations describing the Michel flow. We work in spherical polar coordinates $(t,r,\theta,\phi)$ and assume gravitational units with $c = G = 1$.  A Schwarzschild black hole with mass $m$ is described by the metric
\[ g = -\left( 1 - \frac{2m}{r} \right) dt^2 + \left(1 - \frac{2m}{r} \right)^{-1} dr^2  + r^2 (d\theta^2 + \sin^2 \theta d\phi^2). \]
The accretion flow is governed by the conservation laws
\begin{equation}
\label{aaa7}
\nabla_\mu (\rho u^\mu) = 0, \quad \nabla_\mu T^{\mu \nu} = 0,
\end{equation}
where $\rho$ is the baryonic (rest-mass) density of the gas, $u^\mu$ is its four-velocity, and
\[ T^{\mu \nu} = \rho h u^\mu u^\nu + p g^{\mu \nu} \]
denotes the energy-momentum tensor of the perfect-fluid. Here $h$ and $p$ are the specific enthaply and the pressure, respectively. In the following we restrict ourselves to barotropic equations of state, for which there is a functional relation $p = p(\rho)$, $h = h(\rho)$.\footnote{As shown in~\cite{eCoS15a}, for Michel accretion this condition follows automatically from the assumptions that the accretion flow is smooth, spherically symmetric and steady-state.}

Michel accretion solutions are obtained assuming that the flow of the gas is purely radial (the only non-vanishing components of the four-velocity are $u^t$ and $u^r$), spherically symmetric and stationary (all functions depend on the areal radius $r$ only). In this case Eqs.\ (\ref{aaa7}) can be integrated yielding
\begin{equation}
\label{aaa8}
\rho r^2 u^r = \frac{j_n}{4\pi}, \quad h \sqrt{1 - \frac{2m}{r} + (u^r)^2} = L,
\end{equation}
where $j_n$ and $L > 0$ are constants representing, respectively, the particle flux and the ratio between the energy and particle fluxes through a sphere of constant $r$. Assuming that $\rho > 0$, we have $j_n < 0$ for $u^r < 0$ (this situation corresponds to accretion) and $j_n > 0$ for $u^r > 0$ (the gas is flowing outwards; this situation is sometimes referred to as `wind').

In the following sections we work in dimensionless variables. We define $x = r/(2m)$, and $z = \rho/\rho_0$, where $\rho_0$ is a reference rest-mass density. In the main part of this paper we assume a polytropic equation of state of the form $p = K\rho^\gamma$, where $K$ and $\gamma$ are constants. In this case the specific enthalpy (per unit mass) can be expressed as
\[ h = 1 + \frac{\gamma}{\gamma - 1} K \rho^{\gamma - 1} = 1 + \frac{\gamma}{\gamma - 1} K \rho_0^{\gamma - 1} z^{\gamma - 1}. \]
Setting $\frac{\gamma}{\gamma - 1}K \rho_0^{\gamma - 1} = 1$, we get
\[ h =: f(z) = 1 + z^{\gamma - 1}. \]

Many of the results discussed in this paper are also valid for more general barotropic equations of state, provided they satisfy the following reasonable assumptions on the function $f: [0,\infty]\to \Real$: First, we require $f$ to be twice differentiable, positive and monotonously increasing, such that the sound speed
$$
\nu(z) := \sqrt{ \frac{z}{f(z)}\frac{df}{dz}(z)Ê}
$$
is a well-defined differentiable function of $z > 0$. Further, we assume:
\begin{enumerate}
\item[(i)] $0 < \nu(z) < 1$ for all $z > 0$,
\item[(ii)] $f(z)\to 1$ and $\nu(z)\to 0$ as $z\to 0$,
\item[(iii)] $f(z)\to \infty$ as $z\to \infty$.
\end{enumerate}
Condition (i) restricts the equations of state to those having subluminal sound speeds. Condition (ii) implies that the internal energy of the gas vanishes in the limit of vanishing density, while (iii) means that it diverges as $z\to\infty$. Note that the polytropic equation of state satisfies all three conditions provided $1 < \gamma\leq 2$. In the following, we will require the satisfaction of conditions (i--iii), and when we specialize to the polytropic case, this will be stated explicitly.

In terms of the above dimensionless variables, Eqs.\ (\ref{aaa8}) can be written as
\begin{eqnarray}
\label{fff1}
x^2 z u &=& \mu  = \mathrm{const},\\
f(z)^2\left( 1 - \frac{1}{x} + u^2 \right) &=& L^2 = \mathrm{const},
\label{fff2}
\end{eqnarray}
where for simplicity we denoted $u := u^r$, and $\mu := j_n/(16\pi\rho_0 m^2)$. The same notation is also used in\ \cite{eCoS15a}.

%%%%%%%%%%%%%%%%%%%%%%%%%%%%%%%%%%%%%%%%%%%%%
\section{Michel accretion as a Hamiltonian dynamical system}
\label{sec_dyn_system}
%%%%%%%%%%%%%%%%%%%%%%%%%%%%%%%%%%%%%%%%%%%%%

For what follows, it is convenient to introduce a (fictitious) dynamical system, whose orbits consist of graphs of solutions of system (\ref{fff1}--\ref{fff2}). Such a dynamical system can be defined conveniently either in terms of $x$ and $u$ (as in~\cite{pM15}), or in terms of $x$ and $z$. The latter convention was adopted in \cite{eCoS15a}, and we use it here as well.

Differentiating Eqs.\ (\ref{fff1}) and (\ref{fff2}) with respect to $x$ one obtains
\[ \frac{dz}{dx} = \frac{\frac{1}{x}\left( 4 u^2  - \frac{1}{x} \right)}{\frac{2}{z} \left[ \nu(z)^2 \left(1 - \frac{1}{x} + u^2 \right) - u^2 \right]}, \]
where we have used the relation $df/dz(z) = f(z)\nu(z)^2/z$. Then, choosing a parameter $l$ so that
\begin{equation}
\label{fff3}
\frac{dx}{dl} = \frac{2f(z)^2}{z} \left[ \nu(z)^2 \left(1 - \frac{1}{x} + u^2 \right) - u^2 \right],
\end{equation}
one can write
\begin{equation}
\label{fff4}
\frac{dz}{dl} = \frac{f(z)^2}{x}\left( 4 u^2  - \frac{1}{x} \right).
\end{equation}
Equations (\ref{fff3}) and (\ref{fff4}) define a two-dimensional dynamical system, provided that the terms containing $u^2$ can be reexpressed in terms of $x$ and $z$. This can be achieved by using either Eq.\ (\ref{fff1}) or Eq.\ (\ref{fff2}). In the first case one obtains a dynamical system which depends on the parameter $\mu$. Hence, all orbits correspond to the same value of $\mu$, but, as we will see, they have different values of $L$ associated to them. In the second case, one obtains a dynamical system with $L$ being a parameter. Accordingly, all orbits correspond to the same value of $L$, and they are associated with different values of $\mu$. In both cases the resulting dynamical system is Hamiltonian.

The first option was used in~\cite{eCoS15a} and corresponds to the choice
\[ F_\mu (x,z) :=  \left. f(z)^2 \left( 1 - \frac{1}{x} + u^2 \right) \right|_{u = \frac{\mu}{x^2 z}} = L^2 \]
for the Hamiltonian, with associated equations of motion
\begin{eqnarray}
\label{fff5}
\frac{dx}{dl} & = & \frac{\partial F_\mu}{\partial z}(x,z) = \left. \frac{2 f(z)^2}{z} \left[ \nu(z)^2 \left(1 - \frac{1}{x} + u^2 \right) - u^2 \right] \right|_{u = \frac{\mu}{x^2 z}}, \\
\frac{dz}{dl} & = & - \frac{\partial F_\mu}{\partial x}(x,z) = \left. \frac{f(z)^2}{x} \left(4 u^2 - \frac{1}{x} \right) \right|_{u = \frac{\mu}{x^2 z}},
\label{fff6}
\end{eqnarray}
where $L$ parametrizes the level sets of the Hamiltonian $F_\mu$, and all the orbits have the same value of $\mu$, as explained above.

The second possibility, which we find more convenient for the purpose of the present article, is to assume that the value of $L$ is fixed in the entire phase portrait. Accordingly, we define a new Hamiltonian as
\begin{equation}
H_L(x,z) := x^4 z^2\left( \frac{L^2}{f(z)^2} - 1 + \frac{1}{x} \right) = \mu^2.
\label{Eq:HLDef}
\end{equation}
The corresponding dynamical system reads now
\begin{eqnarray}
\label{fff7}
\frac{dx}{d\tilde l} & = & \frac{\partial H_L}{\partial z}(x,z) = 2x^4 z\left[ \frac{L^2}{f(z)^2}(1 - \nu(z)^2) - 1 + \frac{1}{x} \right], \\
\frac{dz}{d\tilde l} & = & - \frac{\partial H_L}{\partial x}(x,z) = - x^3 z^2\left( \frac{4L^2}{f(z)^2} - 4 + \frac{3}{x} \right).
\label{fff8}
\end{eqnarray}
In this description, it is $\mu$ that parametrizes the level sets of the Hamiltonian, while all the orbits have the same value of $L$. The above system is equivalent  to the one given by Eqs.\ (\ref{fff3}) and (\ref{fff4}) with the substitution
\[ u^2 = \frac{L^2}{f(z)^2} - 1 + \frac{1}{x}, \]
up to the `time' reparametrization of the orbits (a replacement of $l$ by a new parameter $\tilde l$). The general relation between two functionally related Hamiltonians and the corresponding dynamical system is discussed in the Appendix.

Note that in the second formulation the value of the Hamiltonian is directly related to the mass accretion rate $\mu$. As we will show below, the value of the parameter $L$ plays a decisive role in the nature of the transonic flow solutions. In order to illustrate this fact, suppose there exists a solution $z(x)$ of $H_L(x,z(x)) = \mu^2$ which extends all the way to the asymptotic region $x\to \infty$ and has $z_\infty = \lim\limits_{x\to \infty} z(x) > 0$. Then, it follows from Eq.~(\ref{Eq:HLDef}) that $f(z_\infty) = L$. Since $f\geq 1$, this means that a necessary condition for a globally-defined solution to exist is that $L\geq 1$.

It should be stressed that each orbit (or a solution of the accretion problem) can be equivalently represented on phase portraits of systems (\ref{fff5}--\ref{fff6}) or (\ref{fff7}--\ref{fff8}). On the other hand, the phase portraits of systems (\ref{fff5}--\ref{fff6}) and (\ref{fff7}--\ref{fff8}) are composed of different families of orbits; thus in some cases they can look differently.

In the next two sections we work with system~(\ref{fff7}--\ref{fff8}), which allows for a more convenient characterization of both critical points and the types of the orbits. For clarity, we decided to omit an analogous discussion concerning system~(\ref{fff5}--\ref{fff6}), except for a few remarks that we make at the end of Sec.\ \ref{Sec:Global}.

%%%%%%%%%%%%%%%%%%%%%%%%%%%%%%%%%%%%%%%%%%%%
\section{Critical points}
\label{Sec:Crit}
%%%%%%%%%%%%%%%%%%%%%%%%%%%%%%%%%%%%%%%%%%%%

In this section we analyze the critical points of the Hamiltonian system~(\ref{fff7}--\ref{fff8}), which are determined by the zeros of the gradient of $H_L$. Since
\begin{eqnarray}
\frac{\partial H_L}{\partial x}(x,z) &=& x^3 z^2\left( \frac{4L^2}{f(z)^2} - 4 + \frac{3}{x} \right),
\label{Eq:HLx}\\
\frac{\partial H_L}{\partial z}(x,z) &=& 2x^4 z\left[ \frac{L^2}{f(z)^2}(1 - \nu(z)^2) - 1 + \frac{1}{x} \right],
\label{Eq:HLz}
\end{eqnarray}
we find the conditions
\begin{equation}
x = \frac{3}{4} + \frac{1}{4\nu(z)^2},\qquad
\frac{f(z)^2}{1 + 3\nu(z)^2} = L^2
\label{Eq:xzcrit}
\end{equation}
for a critical point. Since $0\leq\nu(z) < 1$, it follows that any critical point $(x,z)$ must lie outside the event horizon.

The number of critical points depends on the behavior of the function ${\cal L}: [0,\infty) \to \Real$ defined by
\begin{equation}
{\cal L}(z) := \frac{f(z)^2}{1 + 3\nu(z)^2},\qquad z\geq 0.
\label{Eq:LDef}
\end{equation}
Note that our assumptions on $f(z)$ and $\nu(z)$ imply that ${\cal L}(0) = 1$ and $\lim\limits_{z\to\infty} {\cal L}(z) = \infty$, and thus there always exists at least one critical point when $L^2 > 1$. If ${\cal L}$ is strictly monotonous, there is precisely one critical point for $L^2 > 1$ and none for $L^2\leq 1$. The following lemma describes the number of critical points as a function of $L$ for the particular case of the polytropic equation of state.

\begin{figure}[htp]
\begin{center}
\includegraphics[width=0.45\textwidth]{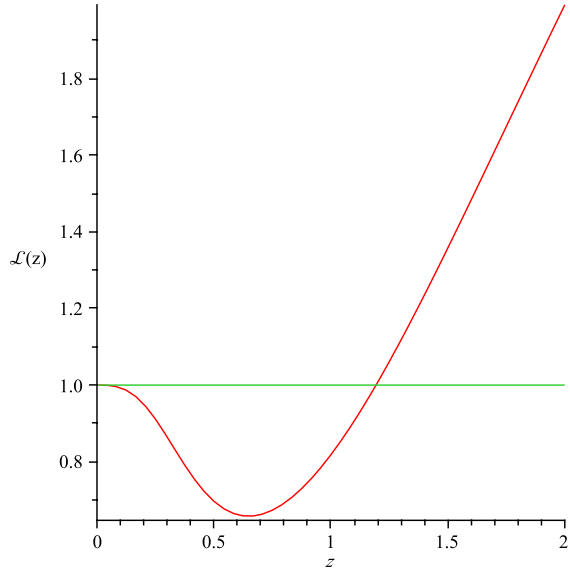}
\end{center}
\caption{\label{Fig:LBehavior} Typical qualitative behavior of the function ${\cal L}(z)$, determining the critical points of the system. In this case, there are two critical points for $L_{min}^2 < L^2 < 1$ and one for $L^2 > 1$.}
\end{figure}

\begin{lemma}
\label{Lem:Crit}
Consider a polytropic equation of state, for which
$$
f(z) = 1 + z^{\gamma-1},\qquad
\nu(z) = \sqrt{(\gamma-1)\frac{z^{\gamma-1}}{1 + z^{\gamma-1}}},\qquad z \geq 0,
$$
with adiabatic index $1 < \gamma \leq 2$. Let
$$
L_{min} := \frac{3}{2}\frac{\gamma-1}{\left(\gamma - \frac{2}{3} \right)^{3/2}}.
$$
Then, system~(\ref{Eq:HLDef}) has a unique critical point for each $L^2 > 1$. When $1 < \gamma\leq 5/3$, there are no critical points for $L^2 < 1$, while for $5/3 < \gamma\leq 2$ there are no critical points for $L^2 < L_{min}^2$ and precisely two critical points for each $L_{min}^2 < L^2 < 1$.
\end{lemma}

\proof Setting $y := \nu(z)^2$, one can rewrite the condition ${\cal L}(z) = L^2$ as
\begin{equation}
G(y) := \frac{1}{1 + 3y}\frac{1}{\left( 1 - \frac{y}{\gamma-1} \right)^2} = L^2.
\label{Eq:GDef}
\end{equation}
Note that $y = \nu(z)^2$ lies in the interval $[0,\gamma-1)$. The function $G: [0,\gamma-1)\to\Real$ defined above satisfies $G(0) = 1$ and $\lim\limits_{y\to \gamma-1} G(y) = \infty$, and it has derivative
$$
\frac{dG}{dy}(y) = \frac{1}{(1 + 3y)^2}\frac{1}{\left( 1 - \frac{y}{\gamma-1} \right)^3}
\frac{9y - (3\gamma - 5)}{\gamma-1}.
$$
When $1 < \gamma \leq 5/3$ it follows that $G$ is strictly monotonically increasing, and hence there exists a unique critical point for each $L^2 > 1$. In contrast to this, when $5/3 < \gamma\leq 2$, $G$ has a global minimum at $y = y_m := (3\gamma-5)/9$, where $G(y_m) = L_{min}^2 < 1$, and the Lemma follows.
\qed

{\bf Remarks}:
\begin{enumerate}
\item $L_{min}^2$ decreases monotonically from $1$ to $3^5/4^4 = 243/256$, as $\gamma$ increases from $5/3$ to $2$.\\

\item It is easy to verify that when $5/3 < \gamma \leq 2$, the function $G(y)$ crosses the value $L^2 = 1$ at $y = 0$ and
$$
y = y_* := \gamma - 1 - \frac{1}{6}\left[ 1 + \sqrt{12\gamma-11} \right].
$$
Therefore, when $L_{min}^2 < L^2 < 1$, the two critical points $(x_1,z_1)$ and $(x_2,z_2)$ satisfy
$$
0 < y_1 < y_m,\qquad
y_m < y_2 < y_*,
$$
and when $L^2 > 1$ the unique critical point $(x_c,z_c)$ lies in the interval $y_* < y_c < y_0 := \gamma-1$. Since $x = 3/4 + 1/(4y)$ the corresponding intervals for the location of the critical point are
$$
x_1 > x_m,\qquad x_m > x_2 > x_*,
$$
and $x_* > x_c > x_0$, where
$$
x_0 := \frac{1}{4}\frac{3\gamma-2}{\gamma-1},\qquad
x_* := \frac{3}{4} + \frac{1}{4y_*},\qquad
x_m := \frac{1}{4}\frac{3\gamma-2}{\gamma - 5/3}.
$$
It follows from the behavior of the function $G(y)$ that as $L^2$ increases from $L_{min}^2$ to $1$, $y_1$ decreases from $y_m$ to $0$ while $y_2$ increases from $y_m$ to $y_*$. Correspondingly, $x_1$ increases from $x_m$ to $\infty$, while $x_2$ decreases from $x_m$ to $x_*$. Furthermore, as $L^2$ increases from $1$ to $\infty$, $y_c$ increases from $y_*$ to $\gamma - 1$, and $x_c$ decreases from $x_*$ to $x_0$.

\item Explicit expressions for the values of $y_1$ and $y_2$ can be given using Cardano's formulae. For $5/3 < \gamma < 2$ and $L_{min}^2 < L^2 < 1$, one has
\begin{eqnarray}
y_1 & = & \frac{1}{9} \left\{ 6 \gamma - 7 - 2(3 \gamma - 2) \cos \left[ \frac{\pi}{3} - \frac{1}{3} \mathrm{arccos} \left( \frac{2 L_{min}^2}{L^2} - 1 \right) \right] \right\}, \\
y_2 & = & \frac{1}{9} \left\{ 6 \gamma - 7 - 2(3 \gamma - 2) \cos \left[ \frac{\pi}{3} + \frac{1}{3} \mathrm{arccos} \left( \frac{2 L_{min}^2}{L^2} - 1 \right) \right] \right\}. \label{cardanoy2}
\end{eqnarray}
For $L^2 > 1$, the unique critical point corresponds to the square of the speed of sound given by Eq.\ (\ref{cardanoy2}) (formula for $y_2$).
\end{enumerate}

Next, we analyze whether the critical points are saddle points or extrema of the Hamiltonian $H_L$. For this, we evaluate the Hessian of $H_L$ at a critical point $(x_c,z_c)$:
\begin{equation}
D^2 H_L(x_c,z_c) = \left. \left( \begin{array}{cc}
\frac{\partial^2 H_L}{\partial x^2} & \frac{\partial^2 H_L}{\partial x\partial z} \\
\frac{\partial^2 H_L}{\partial z\partial x} & \frac{\partial^2 H_L}{\partial z^2}
\end{array} \right) \right|_{(x_c,z_c)}
 = -x_c^3\left( \begin{array}{cc}
3\frac{z_c^2}{x_c^2} & 2\frac{z_c}{x_c} \\
2\frac{z_c}{x_c}  & 1 - \nu_c^2 + w_c
\end{array} \right),
\end{equation}
where we have defined
$$
\nu_c := \nu(z_c),\qquad
w_c := \frac{\partial\log\nu}{\partial\log z}(z_c).
$$
The determinant of the Hessian is
\begin{equation}
\det\left[ D^2 H_L(x_c,z_c) \right] = -x_c^4 z_c^2(1 + 3\nu_c^2 - 3w_c),
\end{equation}
and hence the critical point is a saddle point, if $1 + 3\nu_c^2 - 3w_c > 0$, and a local extremum, if $1 + 3\nu_c^2 - 3w_c < 0$.

As it turns out, the sign of the determinant is related to the sign of the slope of the function ${\cal L}$ defined in Eq.~(\ref{Eq:LDef}) at the critical point. Indeed, a short calculation reveals that
\begin{equation}
\frac{d}{dz}{\cal L}(z_c) = \frac{L^2}{2 x_c z_c}(1 + 3\nu_c^2 - 3w_c)
 = -\frac{L^2}{2x_c^5 z_c^3}\det\left[ D^2 H_L(x_c,z_c) \right].
\end{equation}
Therefore, the critical point is a saddle whenever the function ${\cal L}(z)$ crosses the value $L^2$ from below, and an extremum whenever the function ${\cal L}(z)$ crosses $L^2$ from above. The next lemma describes the situation for a polytropic fluid and is a direct consequence of Lemma~\ref{Lem:Crit} and the above remarks.

\begin{lemma}
\label{Lem:Crit2}
For the particular case of the polytropic equation of state with adiabatic index $1 < \gamma\leq 2$, we have the following:
\begin{enumerate}
\item[(a)] For $L^2 > 1$, the unique critical point is a saddle point lying in the interval $x_c = x_{saddle}\in (x_0,\infty)$.
\item[(b)] When $5/3 < \gamma \leq 2$ and $L_{min}^2 < L^2 < 1$ there is one saddle critical point lying in the interval $x_2 = x_{saddle}\in (x_*,x_m)$ and one center critical point lying in the interval $x_1 = x_{center}\in (x_m,\infty)$. As $L^2$ increases from $L_{min}^2$ to $1$, $x_1 = x_{center}$ increases from $x_m$ to $\infty$, and $x_2 = x_{saddle}$ decreases from $x_m$ to $x_*$. 
\end{enumerate}
\end{lemma}

{\bf Remark}: In fact, the statement of the lemma holds true as long as the function ${\cal L}(z)$ defined in Eq.~(\ref{Eq:LDef}) has a qualitative behavior similar to the one shown in Fig.~\ref{Fig:LBehavior}.

For later use we will also need the slopes of the two level curves of $H_L$ crossing a critical saddle point. For this, let $(x(l),z(l))$ be such a level curve. Differentiating the relation
$$
H_L(x(l),z(l)) = \mu^2
$$
twice with respect to the parameter $l$ and evaluating the result at a saddle point $(x_c,z_c)$ we obtain the quadratic equation
$$
A + 2B \frac{dz}{dx}(x_c) + C \left[ \frac{dz}{dx}(x_c) \right]^2 = 0
$$
for the slope $dz/dx(x_c)$ of the curve at $x = x_c$, where
\begin{equation}
A := \frac{\partial^2 H_L}{\partial x^2}(x_c,z_c) = -3x_c z_c^2,\quad
B := \frac{\partial^2 H_L}{\partial x\partial z}(x_c,z_c) = -2x_c^2 z_c,\quad
C := \frac{\partial^2 H_L}{\partial z^2}(x_c,z_c) = -x_c^3(1 - \nu_c^2 + w_c).
\label{Eq:DefABC}
\end{equation}
Therefore, for $C < 0$, the slope of the two level curves $\Gamma_\pm(x_c,z_c)$ through $(x_c,z_c)$ is given by
\begin{equation}
\frac{dz_{\pm}}{dx}(x_c) = \frac{-B \mp \sqrt{B^2 - AC}}{C} 
 = \frac{z_c}{x_c}\frac{-2\pm \sqrt{1 + 3\nu_c^2 - 3w_c}}{1 - \nu_c^2 + w_c}.
\label{Eq:GammaPMSlope}
\end{equation}

%%%%%%%%%%%%%%%%%%%%%%%%%%%%%%%%%%%%%%%%%%%%
\section{Global extensions}
\label{Sec:Global}
%%%%%%%%%%%%%%%%%%%%%%%%%%%%%%%%%%%%%%%%%%%%

The main goal of this section is to analyze the global behavior of the phase flow defined by the Hamiltonian system~(\ref{Eq:HLDef}). We are particularly interested in the global structure of the local one-dimensional stable ($\Gamma_-$) and unstable  ($\Gamma_+$) manifolds associated with the saddle critical point.

In~\cite{eCoS15a} two of us treated the case of a perfect fluid accreted by a static black hole satisfying certain assumptions. In particular, the results in~\cite{eCoS15a} cover the case of a polytropic fluid with adiabatic index $1 < \gamma \leq 5/3$ accreted by a Schwarzschild black hole, and it was proven that in this case the unstable manifold $\Gamma_+$ associated with the saddle critical point extends all the way from the event horizon to the asymptotic region $x\to\infty$, where the particle density converges to a finite, positive value.

In this section, we discuss the complementary case, that is, the case of a polytropic fluid with adiabatic index $5/3 < \gamma\leq 2$. Our main result shows that when $L^2 > 1$ (the case where there exists a unique critical saddle point) the extension of $\Gamma_-$ has exactly the same qualitative properties as in the case $1 < \gamma\leq 5/3$ and describes a global accretion flow extending from the horizon to the asymptotic region. In contrast to this, when $L_{min}^2 < L^2 < 1$, we prove that $\Gamma_-$ and $\Gamma_+$ connect to each other and form a homoclinic orbit.

We base our global analysis on the two curves $c_1$, $c_2$ in phase space $\Omega := \{ (x,z) : x > 0, z > 0 \}$, corresponding to those points $(x,z)$ with vanishing partial derivative of $H_L$ with respect to $x$ and $z$, respectively.\footnote{Note that the curve $c_2$ has the following physical interpretation: it divides $\Omega$ into a region (above $c_2$) where the flow's radial velocity,
$$
v = -\sqrt{1 - \frac{f(z)^2}{L^2}\left(1 - \frac{1}{x} \right)},
$$
as measured by static observers is subsonic and a region (below $c_2$) where this radial velocity is supersonic.} These curves can be parametrized as follows (cf. Eqs.~(\ref{Eq:HLx},\ref{Eq:HLz})):
\begin{eqnarray}
c_1 &:& x_1(z) = \frac{3}{4}\left[ 1 - \frac{L^2}{f(z)^2} \right]^{-1},\qquad
f(z) > L,
\label{Eq:Gamma1Def}\\
c_2 &:& x_2(z) = \left[ 1 - \frac{L^2}{f(z)^2}(1 - \nu(z)^2) \right]^{-1},\qquad
f(z) > L\sqrt{1 - \nu(z)^2}.
\label{Eq:Gamma2Def}
\end{eqnarray}
By definition, the curves $c_1$ and $c_2$ intersect each other precisely at the critical points of the system, and further they divide the phase space into different regions, the components of the Hamiltonian vector field associated with $H_L$ having a fixed sign in each of these regions.

Before we focus our attention on the polytropic case, let us make a few general remarks about the qualitative behavior of the curves $c_1$ and $c_2$. First, a short calculation reveals that
$$
\frac{dx_1}{dz}(z) = -\frac{8L^2\nu(z)^2}{3zf(z)^2} x_1(z)^2 < 0,
\label{Eq:dx1dz}
$$
and
$$
\frac{dx_2}{dz}(z) = -\frac{2L^2\nu(z)^2}{z f(z)^2} x_2(z)^2\left[ 1 - \nu(z)^2 + w(z) \right],
$$
which is negative as long as $1-\nu^2 + w > 0$.\footnote{Note that since $\nu^2 < 1$, this condition is automatically satisfied, if $w\geq 0$. In particular, it is satisfied in the polytropic case as long as $1 < \gamma \leq 2$.} Consequently, both functions $x_1(z)$ and $x_2(z)$ are monotonically decreasing in $z$. Furthermore, as $z\to \infty$, our general conditions (i) and (iii) imply that
$$
\lim\limits_{z\to\infty} x_1(z) = \frac{3}{4}
 < \lim\limits_{z\to\infty} x_2(z) = 1.
$$
When $z\to 0$, we have $f(z)\to 1$ and $\nu(z)\to 0$ by assumption (ii), and thus the conditions $f(z) > L$ and $f(z) > L\sqrt{1 - \nu(z)^2}$ are only satisfied for all $z\geq 0$, if $L^2 < 1$. In this case
$$
\lim\limits_{z\to 0} x_1(z) = \frac{3}{4}\frac{1}{1 - L^2} 
 < \lim\limits_{z\to 0} x_2(z) = \frac{1}{1 - L^2}.
$$
However, when $L > 1$, there are minimal values $\tilde{z}_1 > 0$ and $\tilde{z}_2 > 0$ such that $f(\tilde{z}_1) = L$ and $f(z) > L$ for all $z > \tilde{z}_1$, while $f(\tilde{z}_2) = L\sqrt{1 - \nu(\tilde{z}_2)^2}$ and $f(z) > L\sqrt{1 - \nu(z)^2}$ for all $z > \tilde{z}_2$. In this case, both $x_1(z)$ and $x_2(z)$ tend to infinity as $z$ decreases to $\tilde{z}_1$ and $\tilde{z}_2$, respectively.

Finally, let us compare the slopes of the curves $c_{1,2}$ to those of the level curves $\Gamma_\pm$ of $H_L$ at a critical saddle point. Using the fact that along $c_1$ we have
$$
\frac{\partial H_L}{\partial x}(x_1(z), z) = 0,
$$
differentiation with respect to $z$ and evaluation at the critical point yields
$$
A\frac{dx_1}{dz}(z_c) + B  = 0,
$$
where the coefficients $A,B$ are defined in Eq.~(\ref{Eq:DefABC}). Consequently, we find that
$$
\left( \frac{dz}{dx} \right)_1(x_c) = -\frac{A}{B},
$$
where the subindex $1$ means that the differentiation is taken along the curve $c_1$. Similarly, the slope of $c_2$ at the critical point is given by
$$
\left( \frac{dz}{dx} \right)_2(x_c) = -\frac{B}{C}.
$$
Comparing these expressions to the ones given in Eq.~(\ref{Eq:GammaPMSlope}) we find, since $A,B,C < 0$ and $B^2 - A C > 0$,
\begin{equation}
\frac{dz_-}{dx}(x_c) < \left( \frac{dz}{dx} \right)_2(x_c) 
 < \left( \frac{dz}{dx} \right)_1(x_c) < \frac{dz_+}{dx}(x_c) < 0.
\end{equation}

In the following, we focus on the polytropic case and analyze the phase space for the two cases $L > 1$ and $L_{min} < L < 1$ separately.

\subsection{Case I: $L > 1$}

\begin{figure}[htp]
\begin{center}
\includegraphics[width=0.6\textwidth]{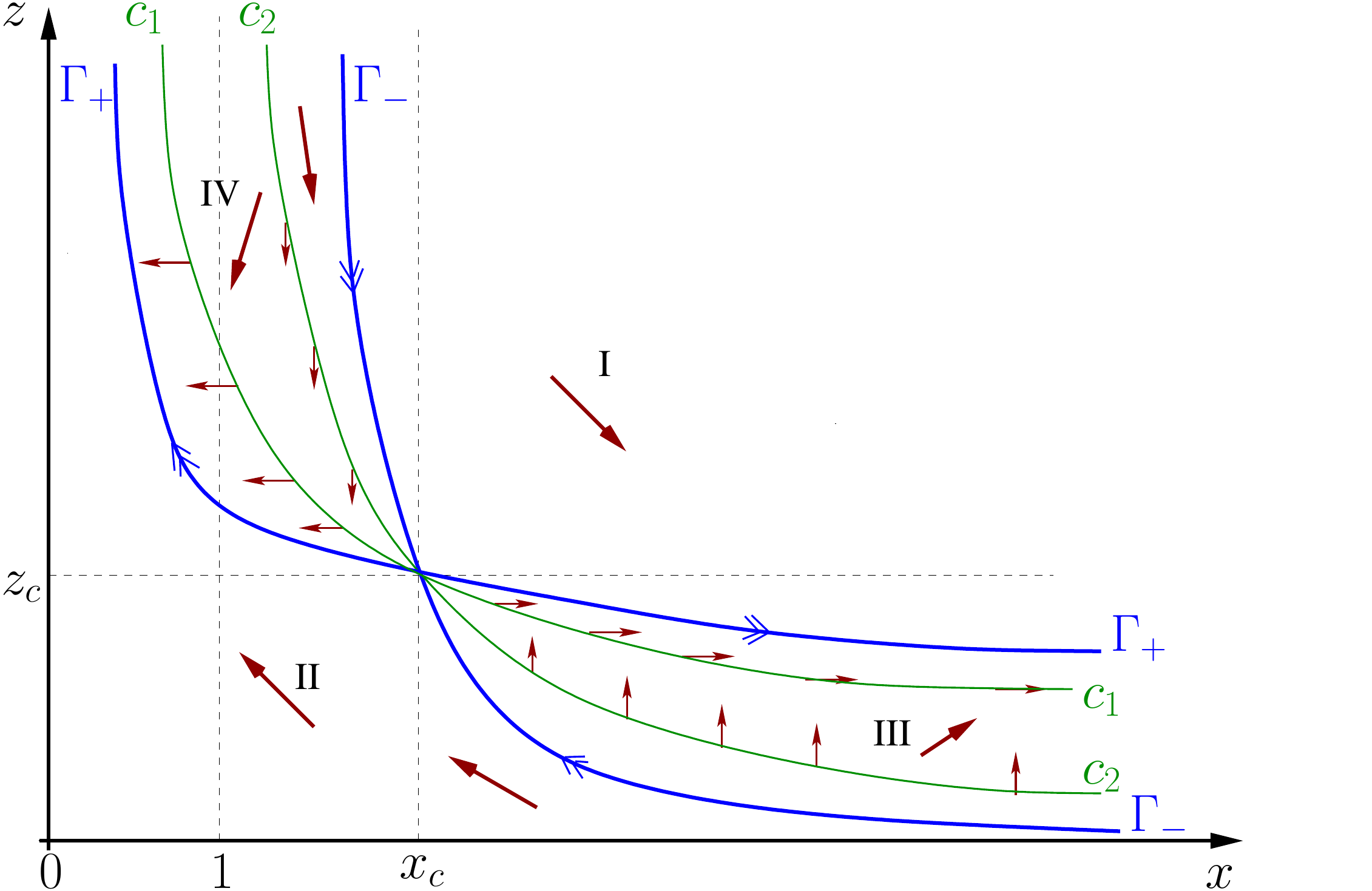}
\end{center}
\caption{\label{Fig:FaseCaseI} A sketch of the phase space, showing the saddle critical point $(x_c,z_c)$ with the associated stable ($\Gamma_-$) and unstable ($\Gamma_+$) manifolds, and the special curves $c_1$ and $c_2$ defined in Eqs.~(\ref{Eq:Gamma1Def}) and (\ref{Eq:Gamma2Def}). Also shown is the direction of the flow in each of the regions and along the curves $c_1$ and $c_2$. The flow's radial velocity measured by static observers is subsonic in the region above the curve $c_2$ and supersonic in the region below $c_2$.}
\end{figure}

According to Lemma~\ref{Lem:Crit2} there is a unique saddle point in this case, and according to the remarks above the qualitative behavior of the curves $c_{1,2}$ and $\Gamma_\pm$ are as shown in Fig.~\ref{Fig:FaseCaseI}.
Following the arguments described in Sec.\ IIIC of~\cite{eCoS15a}, one can show that the unstable manifold $\Gamma_+$ extends to the horizon $x = 1$ on one side of $x_c$, and that it must extend to $x\to \infty$ above the curve $c_1$ on the other side of the critical saddle point. Furthermore, the stable manifold $\Gamma_-$ asymptotes to the horizon $x=1$ with $z\to \infty$ on one side of $x_c$, while it must extend to $x\to \infty$ below the curve $c_2$ on the other side of $x_c$ (see Fig.~\ref{Fig:FaseCaseI}).

\subsection{Case II: $L_{min} < L < 1$ and $5/3 < \gamma \leq 2$}

\begin{figure}[htp]
\begin{center}
\includegraphics[width=0.6\textwidth]{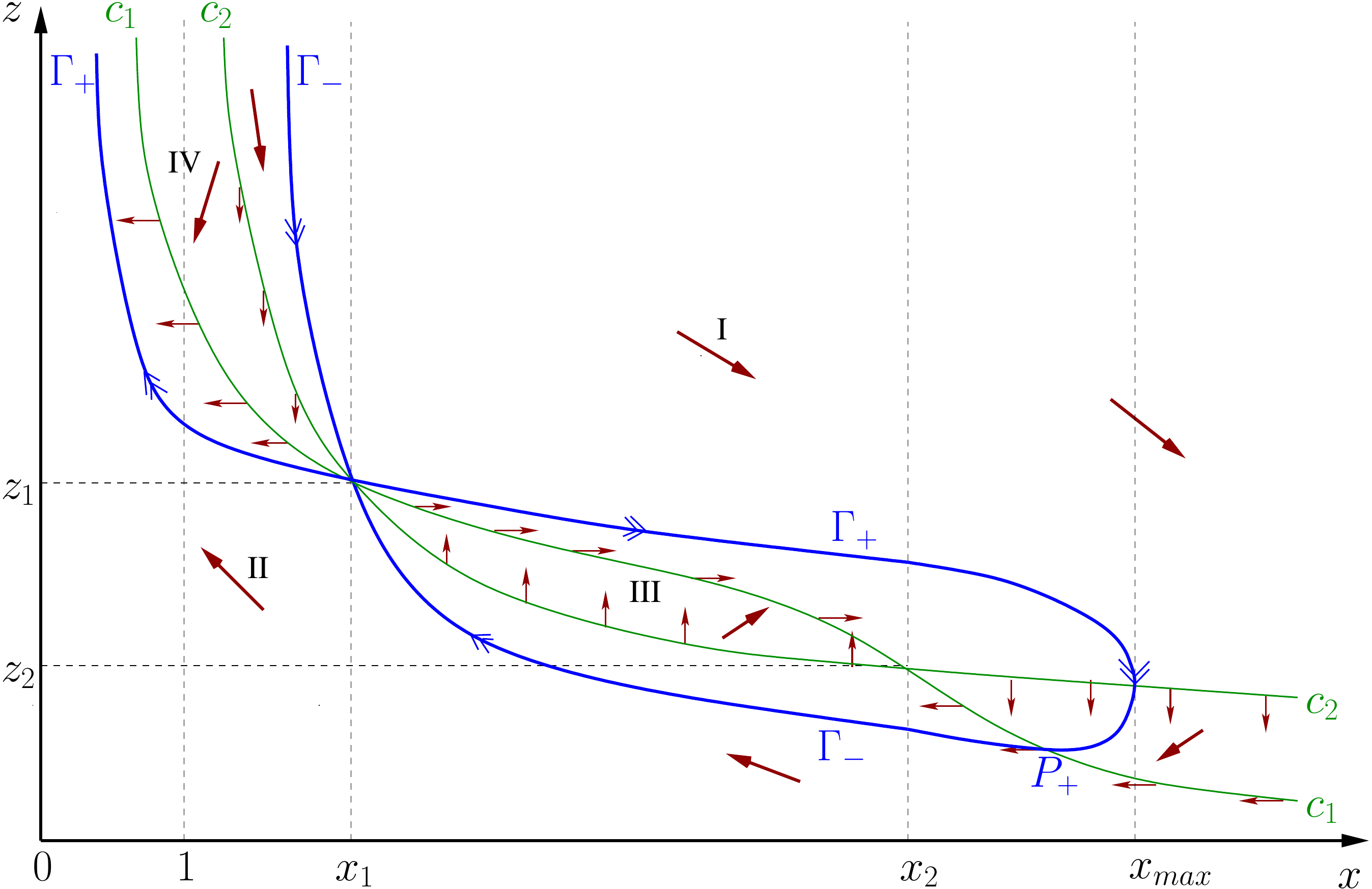}
\end{center}
\caption{\label{Fig:FaseCaseII} A sketch of the phase space, showing the saddle critical point $(x_1,z_1)$ with the associated stable ($\Gamma_-$) and unstable ($\Gamma_+$) manifolds, the center critical point $(x_2,z_2)$, and the curves $c_1$ and $c_2$ defined in Eqs.~(\ref{Eq:Gamma1Def}) and (\ref{Eq:Gamma2Def}). Also shown is the direction of the flow in each of the regions and along the curves $c_1$ and $c_2$. As proven in the text, in this case $\Gamma_+$ and $\Gamma_-$ connect to each other and form a homoclinic orbit.}
\end{figure}

In this case, it follows from Lemmas~\ref{Lem:Crit} and \ref{Lem:Crit2} that there are two critical points, one saddle and one center (see Fig.~\ref{Fig:FaseCaseII} for a sketch of the phase diagram).
The extension of the stable and unstable manifolds $\Gamma_\pm$ towards the horizon is qualitatively the same as in the previous case. Furthermore, $\Gamma_+$ extends to $x = x_2$ above the curve $c_1$ and likewise $\Gamma_-$ extends to $x = x_2$ below the curve $c_2$. However, as we show now, the behavior of the extensions of $\Gamma_\pm$ beyond $x = x_2$ changes radically when compared to the previous case.

First, we notice that $H_L(x,z) = \mu^2 > 0$ implies that
$$
\frac{1}{x} \geq 1 - \frac{L^2}{f(z)^2} \geq 1 - L^2,
$$
which is positive when $L^2 < 1$. Therefore, when $L^2 < 1$, $x$ cannot extend to infinity along the level curves $\Gamma_\pm$ and is bounded by $1/(1-L^2)$. Since along $\Gamma_+$ the coordinate $z$ decreases and $x$ increases as one moves away from the critical saddle point $x_1$, and since $x$ is bounded, it follows that $\Gamma_+$ must intersect the curve $c_2$ at some point $x_{max} > x_2$ (see Fig.~\ref{Fig:FaseCaseII}). Extending further $\Gamma_+$ in the region between $c_2$ and $c_1$ we see that in fact $\Gamma_+$ has to intersect the curve $c_1$ as well. Let us denote this point by $P_+$. Using a similar argument, it follows that $\Gamma_-$ must intersect the curve $c_1$ at some point $P_-$, say. We now claim that $P_+ = P_-$, implying that the two curves $\Gamma_\pm$ connect to each other and form a homoclinic orbit. For this, we use the following lemma.

\begin{lemma}
Consider the function $J(z) := H_L(x_1(z),z)$, $f(z) > L$, which represents the value of the Hamiltonian $H_L$ along the curve $c_1$. Then, $J$ is strictly monotonously increasing on the intervals $(0,z_2)$ and $(z_1,\infty)$ and strictly monotonously decreasing on the interval $(z_1,z_2)$.
\end{lemma}

\proof Differentiating both sides of $J(z) = H_L(x_1(z),z)$ with respect to $z$ we find
$$
\frac{dJ}{dz}(z) = \frac{\partial H_L}{\partial x}(x_1(z),z)\frac{dx_1}{dz}(z) 
 + \frac{\partial H_L}{\partial z}(x_1(z),z).
$$
The first term on the right-hand side is zero because by definition the partial derivative of $H_L$ with respect to $x$ is zero on $c_1$. For the second term, we use Eqs.~(\ref{Eq:HLz},\ref{Eq:Gamma1Def}) and find
$$
\frac{dJ}{dz}(z) = \frac{2}{3} z x_1(z)^4\left[ 1 - \frac{L^2}{{\cal L}(z)} \right],
$$
with ${\cal L}(z)$ the function defined in Eq.~(\ref{Eq:LDef}). As a consequence of the proof of Lemma~\ref{Lem:Crit}, we have ${\cal L}(z) > L^2$ for $0 < z < z_1$ or $z_2 < z < \infty$ and ${\cal L}(z) < L^2$ for $z_1 < z < z_2$, and the lemma follows.
\qed

As a consequence of the previous lemma, the value of $H_L$ must increase as one moves along the curve $c_1$ away from $x = x_2$ with increasing $x$. Since the value of $H_L$ is the same constant along $\Gamma_\pm$, it follows that $P_+ = P_-$ and we have a homoclinic orbit.

Note that the results of this section do not make explicit use of the polytropic equation of state. In fact, they remain valid for any equation of state for which the function ${\cal L}(z)$ defined in Eq.~(\ref{Eq:LDef}) has the same qualitative behavior as the one shown in Fig.~\ref{Fig:LBehavior} and for which the quantity $1 - \nu(z)^2 + w(z)$ is positive.

\subsection{A comment on the properties of system~(\ref{fff5}--\ref{fff6})}

A similar analysis to that presented in the preceding sections can be also carried out for system~(\ref{fff5}--\ref{fff6}) (defined by the Hamiltonian $F_\mu$), but it is difficult to formulate a simple characterization of the behavior of its orbits that would depend on the parameters $\gamma$ and $\mu$ only. The asymptotic behavior of a given orbit depends mainly on the value of $L$ associated with it, and hence a much more clear characterization can be given for system~(\ref{fff7}--\ref{fff8}), defined by the Hamiltonian $H_L$, where $L$ is a parameter.

In addition, contrary to the situation described in the preceding sections (for the system with the Hamiltonian $H_L$), system~(\ref{fff5}--\ref{fff6}) admits phase portraits with two critical points (a saddle and a center-type point) and no homoclinic orbit, as well as phase portraits with a homoclinic orbit present. The former case appears when the value of $F_\mu = L^2$ at the saddle-type critical point is less than one.

%%%%%%%%%%%%%%%%%%%%%%%%%%%%%%%%%%%%%%%%%%%%
\section{Conclusions}
\label{Sec:Conclusions}
%%%%%%%%%%%%%%%%%%%%%%%%%%%%%%%%%%%%%%%%%%%%

The analysis presented in this paper was motivated by a recent discovery (in \cite{Eliana-Master-thesis,eCoS12,eCoS15a}) of homoclinic solutions appearing in the Michel model of relativistic steady accretion. They can be found assuming a Schwarzschild background geometry and a standard polytropic equation of state with adiabatic index $5/3 < \gamma \leq 2$.

The assumption of a polytropic equation of state leads to the equations of the model that are just simple enough to allow for a precise characterization of the critical points and general properties of solutions, although (with the exception of the case $\gamma = 2$, for which Eqs.~(\ref{fff1}) and (\ref{fff2}) lead to a fourth--order polynomial equation for $z$) they are still sufficiently complex not to allow to be solved exactly.

The key concept of the analysis of this paper is the correspondence between the equations describing the accretion flow and a (fictitious) dynamical system, that allows to use the terminology (and methods) originating in the theory of dynamical systems. This dynamical system can be defined in many ways (see the Appendix), and the qualitative behavior of the phase portraits can depend on the parametrization that one uses. Here we chose a version for which the complete characterization of the critical points and the types of the orbits seems to be the simplest. In this formulation there is a homoclinic orbit, whenever two critical points (a saddle and a center-type one) are present in the phase diagram and a unique global solution, which extends from the horizon to the asymptotic region whenever there is a unique critical point.

The physical implication of the fact that a graph of a given solution belongs to a homoclinic orbit is that this particular solution is local---it cannot be extended to arbitrarily large radii. Consequently, such solutions are incompatible with the asymptotic boundary conditions that are imposed on the values of a solution at infinity.

Homoclinic accretion solutions were discovered also for models with a negative cosmological constant and adiabatic indices $\gamma < 5/3$, as well as for isothermal equations of state with the square of the speed of sound less than $1/3$ \cite{pMeMjK13, pMeM13, pM15}. An account for the case including charged black holes can be found in \cite{fF15}. The analysis presented in this paper seems to be by far the most complete, but of course it provides only a partial answer to the question of the subtle interplay between the assumed equation of state, the form of the metric, and the existence of homoclinic solutions. It is however worth noticing that the analysis of Sec.\ \ref{Sec:Global} only weakly depends on the equation of state.

%%%%%%%%%%%%%%%%%%%%%%%%%%%
%%%   ACKNOWLEDGMENTS
%%%%%%%%%%%%%%%%%%%%%%%%%%%

\acknowledgments

PM acknowledges discussions with Piotr Bizo\'{n}, Edward Malec and Zdzis{\l}aw Golda. This research was supported in part by CONACyT Grants No. 238758, by the Polish Ministry of Science and Higher Education grant IP2012~000172, by the NCN grant DEC-2012/06/A/ST2/00397 and by a CIC Grant to Universidad Michoacana.

%%%%%%%%%%%%%%%%%%%%%%%%%%%%%%%%%%%%%%%%%%%
%% APPENDIX
%%%%%%%%%%%%%%%%%%%%%%%%%%%%%%%%%%%%%%%%%%%

\appendix*
\section{}

In this appendix we show that two Hamiltonians related functionally as in Sec.\ \ref{sec_dyn_system} define the same orbits.

Consider two differentiable functions $H_1$ and $H_2$ of two variables $x$ and $z$. Suppose we have the following relation between $x$, $z$, $H_1(x,z)$, and $H_2(x,z)$:
\begin{equation}
\label{app0}
G(x, z, H_1(x,z), H_2(x,z)) = 0,
\end{equation}
with a function $G$ which is partially differentiable in all its arguments and which satisfies
\[ \frac{\partial G}{\partial H_1} \neq 0, \qquad \frac{\partial G}{\partial H_2}\neq 0. \]
For example, for the two Hamiltonians introduced Sec.\ \ref{sec_dyn_system}, we have $H_1(x,z) = F_\mu(x,z)$ and $H_2(x,z) = H_L(x,z)$, and they are functionally related by
\[ G(x,z,H_1,H_2) := x^4 z^2 H_1 - f(z)^2\left[ x^4\left( 1 - \frac{1}{x} \right)z^2 + H_2 \right] = 0. \]

Differentiating Eq.\ (\ref{app0}) with respect to $x$ and $z$ we see that
\begin{equation}
\label{app1}
\frac{\partial G}{\partial x} + \frac{\partial G}{\partial H_1} \frac{\partial H_1}{\partial x} + \frac{\partial G}{\partial H_2} \frac{\partial H_2}{\partial x} = 0, \quad \frac{\partial G}{\partial z} + \frac{\partial G}{\partial H_1} \frac{\partial H_1}{\partial z} + \frac{\partial G}{\partial H_2} \frac{\partial H_2}{\partial z} = 0.
\end{equation}

The two dynamical systems that we consider in Sec.\ \ref{sec_dyn_system} are defined by two Hamiltonians, denoted here by $H_1$ and $H_2$, with the following property: For the system defined by the Hamiltonian $H_1$, the function $H_2(x,z) = \mu^2$ is a constant, and it is treated as a parameter. Conversely, for the system with the Hamiltonian given by $H_2$, the function $H_1(x,z) = L^2$ is a constant parameter.

Consider now an orbit in the system with the Hamiltonian $H_1(x,z)$ and $H_2 = \mathrm{const}$. It is given as a solution of the system of equations
\[ \frac{dx}{dl} = \frac{\partial H_1(x,z)}{\partial z} = - \frac{\frac{\partial G(x,z,H_1(x,z),H_2)}{\partial z}}{\frac{\partial G(x,z,H_1(x,z),H_2)}{\partial H_1}}, \quad \frac{dz}{dl} = -\frac{\partial H_1(x,z)}{\partial x} =  \frac{\frac{\partial G(x,z,H_1(x,z),H_2)}{\partial x}}{\frac{\partial G(x,z,H_1(x,z),H_2)}{\partial H_1}}, \]
where we have used Eq.\ (\ref{app1}). Conversely, for a system defined by the Hamiltonian $H_2(x,z)$ with $H_1 = \mathrm{const}$ one obtains
\[ \frac{dx}{d \tilde l} = \frac{\partial H_2(x,z)}{\partial z} = - \frac{\frac{\partial G(x,z,H_1,H_2(x,z))}{\partial z}}{\frac{\partial G(x,z,H_1,H_2(x,z))}{\partial H_2}},
\quad \frac{dz}{d \tilde l} = -\frac{\partial H_2(x,z)}{\partial x} =  \frac{\frac{\partial G(x,z,H_1,H_2(x,z))}{\partial x}}{\frac{\partial G(x,z,H_1,H_2(x,z))}{\partial H_2}}. \]
Because for an autonomous system, the Hamiltonian is a constant of motion, we easily see that for a given orbit (described in both systems) one has
\[ \frac{dx}{dl} = \frac{\frac{\partial G (x,z,H_1,H_2)}{\partial H_2}}{\frac{\partial G (x,z,H_1,H_2)}{\partial H_1}} \frac{d x}{d \tilde l}, \quad
\frac{dz}{dl} = \frac{\frac{\partial G (x,z,H_1,H_2)}{\partial H_2}}{\frac{\partial G (x,z,H_1,H_2)}{\partial H_1}} \frac{d z}{d \tilde l}. \]
This means that the solutions in both Hamiltonian systems are equivalent, up to a reparametrization of the `time' parameter $l$ or $\tilde l$, as claimed.

The above result does not mean, however, that the phase portraits would look the same in both cases. The reason behind this fact is that, in general, they are composed of different orbits. Systems (\ref{fff5}--\ref{fff6}) and (\ref{fff7}--\ref{fff8}) considered in this paper provide a good illustration of this last fact. While for system (\ref{fff7}--\ref{fff8}) we have already proved that the existence of a center-type critical point implies the existence of a homoclinic orbit, one can have a phase portrait of system (\ref{fff5}--\ref{fff6}) with a center-type critical point and no homoclinic orbit.

\bibliographystyle{unsrt}
\bibliography{refs_accretion}

\begin{thebibliography}{10}

\bibitem{hB52}
H.~Bondi.
\newblock On spherically symmetrical accretion.
\newblock {\em Mon. Not. R. Astron. Soc.}, 112:195--204, 1952.

\bibitem{fM72}
F.C. Michel.
\newblock Accretion of matter by condensed objects.
\newblock {\em Astrophys. Space Sci.}, 15:153--160, 1972.

\bibitem{eM99}
E.~Malec.
\newblock Fluid accretion onto a spherical black hole: Relativistic description
  versus {B}ondi model.
\newblock {\em Phys. Rev. D}, 60:104043, 1999.

\bibitem{jKbKpMeMzS06}
J.~Karkowski, B.~Kinasiewicz, P.~Mach, E.~Malec, and Z.~\'{S}wierczy\'{n}ski.
\newblock Universality and backreaction in a general-relativistic accretion of
  steady fluids.
\newblock {\em Phys. Rev. D}, 73(2):021503(R), 2006.

\bibitem{pMeM08}
P.~Mach and E.~Malec.
\newblock Stability of self-gravitating accreting flows.
\newblock {\em Phys. Rev. D}, 78(12):124016, 2008.

\bibitem{jKeMkRzS09}
J.~Karkowski, E.~Malec, K.~Roszkowski, and Z.~\'{S}wierczy\'{n}ski.
\newblock Transonic and subsonic flows in general relativistic radiation
  hydrodynamics.
\newblock {\em Acta Phys. Pol. B}, 40(2):273--293, 2009.

\bibitem{eMtR10}
E.~Malec and T.~Rembiasz.
\newblock General relativistic versus {N}ewtonian: A universality in
  spherically symmetric radiation hydrodynamics for quasistatic transonic
  accretion flows.
\newblock {\em Phys. Rev. D}, 82(12):124005, 2010.

\bibitem{vDyE11}
V.I. Dokuchaev and Yu.N. Eroshenko.
\newblock Accretion with back reaction.
\newblock {\em Phys. Rev. D}, 84(12):124022, 2012.

\bibitem{jKeM13}
J.~Karkowski and E.~Malec.
\newblock Bondi accretion onto cosmological black holes.
\newblock {\em Phys. Rev. D}, 87:044007, 2013.

\bibitem{eBvDyE11}
E.O. Babichev, V.I. Dokuchaev, and Yu.N. Eroshenko.
\newblock Perfect fluid and scalar field in the {R}eissner-{N}ordstrm metric.
\newblock {\em Journal of Experimental and Theoretical Physics}, 112:784, 2011.

\bibitem{sGsKaRtD07}
S.~Goswami, S.N. Khan, A.K. Ray, and T.K. Das.
\newblock Axisymmetric black hole accretion in the {K}err metric as an
  autonomous dynamical system.
\newblock {\em Mon. Not. R. Astron. Soc.}, 378:1407--1417, 2007.

\bibitem{tDbC12}
T.K. Das and B.~Czerny.
\newblock Hysteresis effects and diagnostics of the shock formation in low
  angular momentum axisymmetric accretion in the {K}err metric.
\newblock {\em New Astronomy}, 17:254, 2012.

\bibitem{eTsMjM12}
J.C.~Miller E.~Tejeda, S.~Mendoza.
\newblock Analytic solutions to the accretion of a rotating finite cloud
  towards a central object~{II}. {S}chwarzschild spacetime.
\newblock {\em Mon. Not. R. Astron. Soc.}, 419:1431, 2012.

\bibitem{eTpTjM13}
J.C.~Miller E.~Tejeda, P.A.~Taylor.
\newblock An analytic toy model for relativistic accretion in {K}err spacetime.
\newblock {\em Mon. Not. R. Astron. Soc.}, 429:925, 2013.

\bibitem{fLmGfG14}
F.D. Lora-Clavijo, M.~Gracia-Linares, and F.S. Guzm\'an.
\newblock Horizon growth of supermassive black hole seeds fed with collisional
  dark matter.
\newblock {\em Mon. Not. R. Astron. Soc.}, 443:2242--2251, 2014.

\bibitem{pMeMjK13}
P.~Mach, E.~Malec, and J.~Karkowski.
\newblock Spherical steady accretion flows: {D}ependence on the cosmological
  constant, exact isothermal solutions, and applications to cosmology.
\newblock {\em Phys. Rev. D}, 88:084056, 2013.

\bibitem{pMeM13}
P.~Mach and E.~Malec.
\newblock Stability of relativistic {B}ondi accretion in
  {S}chwarzschild-(anti-)de {S}itter spacetimes.
\newblock {\em Phys. Rev. D}, 88:084055, 2013.

\bibitem{pM15}
P.~Mach.
\newblock Homoclinic accretion solutions in the {S}chwarzschild-anti-de
  {S}itter spacetime.
\newblock {\em Phys. Rev. D}, 91:084016, 2015.

\bibitem{fF15}
F.~Ficek.
\newblock Bondi-type accretion in the {R}eissner-{N}ordstrom-(anti-)de {S}itter
  spacetime.
\newblock {\em Class. Quantum Grav.}, 32(23):235008, 2015.

\bibitem{Eliana-Master-thesis}
E.~Chaverra.
\newblock {\em Accretion of Matter in Spherical Symmetry (Master Thesis)}.
\newblock UMSNH, Mexico, 2011.

\bibitem{eCoS12}
E.~Chaverra and O.~Sarbach.
\newblock Polytropic spherical accretion flows on {S}chwarzschild black holes.
\newblock {\em AIP Conf.Proc.}, 1473:54--58, 2012.

\bibitem{eCoS15a}
E.~Chaverra and O.~Sarbach.
\newblock Radial accretion flows on static, spherically symmetric black holes.
\newblock {\em Class. Quantum Grav.}, 32:155006, 2015.

\bibitem{mB78}
M.C. Begelman.
\newblock Accretion of $\gamma > 5/3$ gas by a {S}chwarzschild black hole.
\newblock {\em Astron. Astrophys.}, 70:583--584, 1978.

\end{thebibliography}
%%%%%%%%%%%%%%%%%%%%%%%%%%%%%%%%%%%%%%%%%%%%

\end{document}